\documentclass[a4paper,oneside,11pt]{article}
\usepackage{amsfonts,amssymb,bm,amsthm,amsmath,float, mathrsfs}
\usepackage[mathcal]{euscript}
\usepackage{enumerate}
\usepackage{subfig,graphicx}
\usepackage{appendix}
\usepackage{setspace}
\usepackage{fancyhdr}
\usepackage{epstopdf}
\usepackage[english]{babel}
\usepackage[round, sort, numbers, authoryear]{natbib}
\numberwithin{equation}{section}

\newtheorem{definition}{Definition}
\newtheorem{proposition}{Proposition}

\usepackage[usenames, dvipsnames]{color}
\numberwithin{proposition}{section}
\numberwithin{lemma}{section}
\numberwithin{definition}{section}
\numberwithin{remark}{section}
\numberwithin{corollary}{section}
\numberwithin{theorem}{section}

\def\SN{\textsf{SN}}	
\def\sn{\textsf{sn}}	
\def\SND{\textsf{Skew-N}}	
\newcommand\Mark[1]{\textsuperscript#1}
\providecommand{\keywords}[1]{{\textit{Keywords}:} #1}
\providecommand{\myabstract}[1]{{\textit{Abstract}:} #1}

\begin{document}

 \thispagestyle{empty}
 
 \fancypagestyle{empty}{%
 \renewcommand{\headrulewidth}{0pt}
  \fancyhf{}% Clear header/footer
  \fancyhead[L]{Published in {\it Statistica Neerlandica}, 2016, vol. 70, no 4, p. 396-413.}% Your journal/note
}
 %\maketitle
 %% Title and authors
\begingroup
\centering
{\LARGE \bfseries A Skew-Normal Copula-Driven GLMM} \\[1.5em]

\large Kalyan Das\Mark{1}, Mohamad Elmasri\Mark{2} and Arusharka Sen\Mark{3}\\
\it{
\Mark{1}University of Calcutta,\Mark{2}McGill University and \Mark{3}Concordia University}
\endgroup

%% Abstract

\begin{changemargin}{0.25in}{0.25in}

\myabstract{This paper presents a method for fitting a copula-driven generalized linear mixed models. For added flexibility, the skew-normal copula is adopted for fitting. The correlation matrix of the skew-normal copula is used to capture the dependence structure within units, while the fixed and random effects coefficients are estimated through the mean of the copula. For estimation, a Monte Carlo expectation-maximization algorithm is developed. Simulations are shown alongside a real data example from the Framingham Heart Study.
}

\end{changemargin}

\section{Introduction}%
The key component driving the development of linear mixed models is the ability of such models to handle data with correlated observations; a data structure where predictors and response variables are measured at more than one level. Such structure is common with repeated observations as in medical studies, where patient characteristics are measured at several time points, not necessarily the same set for each patient. \cite{Fisher} proposed the addition of a random effects term to the linear model, which introduced heteroscedasticity. As a result, the linear mixed model takes the form
\begin{equation}\label{eqn:LMM} \boldsymbol{ Y_i = X_i \beta + D_ib_i} +{\boldsymbol \epsilon_i \:}, \quad i = 1,\dots, m
\end{equation}
where $\boldsymbol{Y_i}$ is an $(n_i \times 1)$ vector of observed
response variable for sample unit $i$, $i =1, ...., m$. $\boldsymbol
{X_i}$ is an $(n_i \times p)$ fixed effects design matrix with
coefficient $\boldsymbol \beta$ of dimension $(p \times
1)$. $\boldsymbol {D_i}$ is an $(n_i \times q)$ random effects design
matrix with coefficient $\boldsymbol {b_i}$ of dimension $(q \times
1)$, and $\boldsymbol{ \epsilon_i}$ is an $(n_i \times 1)$ vector of
random errors. 
%Note that the terms in $\boldsymbol{D_i}$ represent time invariant or categorical variables that constitute the hierarchical structure of the data.
Inference from linear mixed model becomes slightly more tedious by the introduction of the random coefficient $\boldsymbol{b}_i$. This requires an identifiability assumption of independence between $\boldsymbol{b}_i$ and $\boldsymbol{\epsilon}_i$. A popular modeling assumption is then

\begin{equation}\label{eqn:LMMran}\boldsymbol{ b_i} \stackrel{iid}{\sim} N_q(0,\boldsymbol \Omega_b) , \quad \boldsymbol{\epsilon_i} \stackrel{ind}{\sim} {N_n}_i(0,\boldsymbol{\psi_i}), \end{equation}

\noindent where $\boldsymbol{\Omega=\Omega}(\alpha)$ and $\boldsymbol{\psi_i = \psi_i}(\gamma)$ are 	associated dispersion matrices that capture possible variability among -and within- individuals, parametrized by $\alpha$ and $\gamma$. In many literature reviews, the extra restrictiveness associated with specifying the distribution functions of $\boldsymbol{b_i}$  and $\boldsymbol{\epsilon_i}$ is deemed unnecessary. Thereupon, \cite{Vale05} proposed the use of skew-normal in lieu of the normal distribution for both $\boldsymbol{b_i}$ and $\boldsymbol{\epsilon_i}$, in an attempt to capture any slight departures from normality. Moreover, they have explicitly characterized the likelihood function of the resulting model, and fitted it by the constrained expectation maximization algorithm (CEM). Nevertheless, many researchers discussed other techniques and  models for inference, for instance the use of mixture of normals as in \cite{Verbeke}, semi-parametric models as in \cite{Zhang}, non-parametric or smoothed non-parametric technique in maximum likelihood estimation as in \cite{Newton} and predictive recursion algorithm as in \cite{Tao}. This paper follows the \cite{Vale05} approach by modelling the dependence structure in hierarchical multivariate distributions via a copula-driven generalized linear mixed model.

Given response variables $Y_{ij},\; i= 1, \dots n,\; j= 1, \dots, n_i$, we assume that $\boldsymbol{Y_i}= (Y_{i1}, \dots , Y_{in_i})^\top$ follows an $n_i$-variate distribution with a predefined mean and covariance matrix. We model such distribution by using an $n_i$-variate skew-normal copula $\SN_{n_i}(.)$, where the random effects are integrated in the mean structure of the copula. We chose the covariance matrix $\boldsymbol{\Sigma_i}=\boldsymbol{\Sigma}(\xi_i,\boldsymbol{t_i})$ to be of an autoregressive structure in order to include the time-variant parameters. Formally,

\begin{equation} \label{eqn:yi} \boldsymbol{Y_i|b_i} \sim F_{n_i}(\eta(\boldsymbol{X_i \beta +D_ib_i}) , \boldsymbol{\Sigma}(\xi_i,\boldsymbol{t_i}) ) \end{equation}

\noindent where $ \boldsymbol {X_i,\beta, b_i, D_i}$ as defined in \eqref{eqn:LMM} and \eqref{eqn:LMMran}, $\xi_i$ is the dispersion autoregressive time-variant parameter with respect to $\boldsymbol{t_i} = (t_{i1}, \dots, t_{in_i})$, and $\eta(.)$ is a link function. \(F_{k}(\eta, \boldsymbol{\Sigma}) \) is a \(k\)-variate distribution function with mean \(\eta\) and covariance \(\boldsymbol{\Sigma}\). Moreover, we assume the marginal densities $Y_{ij}|b_i$ are a function of $\{ x_{ij},t_{ij}, \boldsymbol{D_i,b_i,\beta} \}$ via the same link function $\eta$. 

The rest of this paper is organized as follows. Section 2 introduces a specific characterization of the skew-normal distribution and the copula used in this paper. Section 3 introduces the model, and constructs the likelihood using a skew-normal copula within a GLM framework. Section 4 discusses the use of numerical Monte Carlo EM algorithm to estimate parameters. Section 5 illustrates simulation results under different models. Section 6, a real data analysis is performed to illustrate the application of our study. Section 7 ends with a general discussion.

\section {Skew-normal distribution and copula}\label{sec:skew-normalcopula}
For a better understanding, we begin this section with the definition of the multivariate skew-normal distribution considered through this paper.

\begin{definition} An \(n\)-dimensional random vector $\boldsymbol X \in \mathbb R^n$ follows a skew-normal distribution with location vector $\boldsymbol \mu \in \mathbb R^n$, dispersion matrix $\boldsymbol \Sigma$ (a $n \times n$ positive definite matrix) and a skewness vector $\boldsymbol{\lambda} \in \mathbb R^n$, if its density function is given by
\begin{equation} \sn_n(\boldsymbol{ x|\mu ,\Sigma,\lambda}) =2\phi_n(\boldsymbol{x|\mu,\Sigma})\Phi_1(\boldsymbol{\lambda^\intercal \Sigma^{-1/2} (x-\mu)}) , \;\;\; \boldsymbol x \in \mathbb R^n. \label{eqn:mform}\end{equation}
In the univariate case
\begin{equation}  \sn_1(x|\mu,\sigma^2, \lambda) = 2\phi_1(x|\mu, \sigma^2) \Phi_1(\lambda \frac{x-\mu}{\sigma}), \label{eqn:uninorm}\end{equation}
\[ (-\infty < x, \mu < \infty) , \quad \mu, \sigma \in \mathbb R , \quad  0<\sigma < \infty.\]
\end{definition}

Here $\phi_n(.|\boldsymbol{\mu, \Sigma)}$ and $\Phi_n(.|\boldsymbol{\mu, \Sigma})$ denote respectively an \(n\)-variate density and distribution function of a normal random variable with mean vector $\boldsymbol \mu$ and covariance matrix $\boldsymbol \Sigma$ ($\sigma^2$ in the univariate case). This notation is used throughout this paper. A special case is when $\boldsymbol \lambda =0$, which reduces the skew-normal to the normal distribution.

The skew-normal characterization in \eqref{eqn:mform} is attributed to \cite{Genton}, and the one in \eqref{eqn:uninorm} is attributed to \cite{Azza85} and expanded further by \cite{AzzDalle96}. Many authors have proposed different forms. However, for convenience, a variation of the characterization in \eqref{eqn:mform} is the only one used in this paper.

\cite{AzzDalle96} proposed a simplified parametrization of $\boldsymbol \lambda$,
in \eqref{eqn:mform}, in terms of an arbitrary $n \times n$ positive definite matrix $\boldsymbol \Delta$, as
\begin{equation}\boldsymbol{ \lambda = \frac{\Delta^{-1/2} \delta}{\sqrt{1-\delta^\intercal \Delta^{-1} \delta}}} \label{eqn:lambdapar}, \end{equation}
where $\boldsymbol{\delta^\intercal \Delta^{-1} \delta} < 1$ for some $ \boldsymbol \delta \in \mathbb R^n$. This characterization is used later to define the likelihood function. 

\subsection{ Skew-normal copula}\label{sec:skewcopula}
A principal part of constructing the copula is defining the marginal distribution of $Y_{ij}|b_i$. In \eqref{eqn:yi}, denote the marginal distribution and density function of $Y_{ij}|b_i$ by $F(y_{ij}|\theta_{ij})$ and $f(y_{ij}|\theta_{ij})$, where $\boldsymbol{ \theta_i}= (\theta_{i1}, \dots, \theta_{in_i})$ are the parameters of interest.

For the same notations in \eqref{eqn:yi}, conditionally on $b_i$ define 
\begin{equation}\label{eqn:zuni}
\boldsymbol {Z_i} = (Z_{i1} , \dots, Z_{in_i}) ^\intercal \sim \SND_{n_i}(\boldsymbol { D_ib_i , \Sigma_i , \lambda_i}),
$$ where the $j$th marginal is $$
Z_{ij} \sim \SND_1((\boldsymbol{D_ib_i})_j ,1, \lambda^*_{ij}),
\end{equation}

where \(\lambda^*_{ij}\) is the univariate skewness parameter, which is not equivalent to the components of the skewness vector \(\boldsymbol {\lambda_i} = (\lambda_{i1}, \dots,
\lambda_{in_i})^\intercal\), rather it is derived using a linear transformation of the multivariate response variable, see Chapter 5 of \cite{azzalini2013skew} for a detailed review. Note that $(\boldsymbol {D_ib_i})_j$ is the $j$th element of the vector $\boldsymbol {D_ib_i}$ and $\boldsymbol{\Sigma_i}=\boldsymbol{\Sigma}(\xi_i,\boldsymbol{t_i})$ is a correlation matrix, which has all its diagonal elements equal to 1.
 
Since the random number $F(Y_{ij}|\theta_{ij}) \sim$ uniform(0,1), we link the two marginal distributions of $Z_{ij}$ and $Y_{ij}$ in a way that for each observation $y_{ij}$ we have 
\begin{equation} \label{eqn:martoscopula}
z_{ij} = {}\SN^{-1}_1[F(y_{ij}|\theta_{ij})|(\boldsymbol {D_ib_i})_j,1,\lambda^*_{ij}],
\end{equation}
and 
\[ \boldsymbol{z_i} = (z_{i1}, \dots, z_{in_i}) 
= \bigg ({}\SN^{-1}_1[F(y_{i1}|\theta_{i1})|\cdot_{i1}], \dots, {}\SN^{-1}_1[F(y_{in_i}|\theta_{in_i})|\cdot_{in_i}]\bigg ), \]
where \(\SN_k\) is a \(k\)-variate skew-normal distribution function.

For presentation simplicity, $\cdot_{ij} = \{(\boldsymbol{D_ib_i})_j, 1, \lambda^*_{ij} \}$ in the above equation. By the transformation in \eqref{eqn:martoscopula}, we attempt to estimate the joint distribution of $\boldsymbol{Y_i|b_i}$ using a copula as 
\begin{equation} \label{eqn:skewcopuladist}F_{n_i}(\boldsymbol{y_i}  |\boldsymbol{\theta_i})={}\SN_{n_i}(\boldsymbol{z_i}|\boldsymbol { D_ib_i , \Sigma_i , \lambda_i}).
\end{equation}

The corresponding density is then
\begin{equation}\label{eqn:comf} f_{n_i}(\mathbf{y_i} |\boldsymbol{\theta_i}) ={}\sn_{n_i} (\mathbf{z_i} |\boldsymbol { D_ib_i , \Sigma_i , \lambda_i})  \prod_{j=1}^{n_i} \frac{ f(y_{ij}|\theta_{ij})}{ {}\sn_1 (z_{ij}|(\boldsymbol{D_ib_i})_j,1,\lambda^*_{ij} )} .\end{equation}

See \cite{Landsman}  for a good reference on skew elliptical copulas and \cite{SIM:SIM1249} for copula-based longitudinal models.
\section{Log-likelihood function}
Despite the defined copula in \eqref{eqn:skewcopuladist} and \eqref{eqn:comf}, writing down the complete log-likelihood function is still difficult. The skew-normal density in \eqref{eqn:mform} is defined partially by the normal distribution function, noted as $\Phi$. Therefore, we first show that the skew-normal copula in \eqref{eqn:zuni} could be simplified by conditioning on latent random variable with a half-normal distribution. By Proposition 1 and Corollary 1 of \cite{Vale05}, based on a characterization due to \cite{Henze86}, we can rewrite the skew-normal distribution of $\boldsymbol Z_i$ as follows.
\[
\boldsymbol{Z_i} \stackrel{\texttt{d}}{=} \boldsymbol{D_ib_i  + \Sigma_i^{1/2} \delta^*_i }v_i + \boldsymbol{\Sigma_i^{1/2} (I - \delta^*_i {\delta^*_i}^\intercal)^{1/2}X_i}
\]
where "\(\stackrel{\texttt{d}}{=}\)" meaning "distributed as", \(v_i \sim HN_1(0,1)\)(HN = half-normal), \(\boldsymbol{X_i}\sim N_{n_i}(0, \boldsymbol{I})\), \(\boldsymbol{b_i}\sim N_q(0,\boldsymbol{\Omega_b})\) are independent and 

$$ \boldsymbol{\delta_i^* = \frac{\lambda_i}{\sqrt{ 1+\lambda_i^\intercal \lambda_i}} }.$$

In other words, 
\begin{equation}\label{eqn:zi} 
\boldsymbol{Z_i} |v_i ,\boldsymbol{b_i} \sim N_{n_i} ( \boldsymbol{D_ib_i  + \Sigma_i^{1/2} \delta^*_i }v_i , \boldsymbol{\Sigma_i^{1/2} (I - \delta^*_i {\delta^*_i}^\intercal) \Sigma_i^{1/2}}), 
\end{equation}
$$v_i \sim HN_1(0,1) , \quad  \boldsymbol{b_i} \sim N_q( 0, \boldsymbol{\Omega_{b}}).$$

Similarly in the univariate case,
\begin{equation}\label{eqn:multiz} Z_{ij} | v_i , \boldsymbol{b_i} \sim N_1( (\boldsymbol{D_ib_i})_j + \delta_{ij} v_i , 1 - \delta^2_{ij} ) \end{equation}
$$ v_i \sim HN_1(0,1), \quad  \boldsymbol{b_i} \sim N_q( 0, \boldsymbol{\Omega_{b}}),$$
where,
$$ \delta_{ij} = \frac{\lambda^*_{ij}}{ \sqrt{ 1+{\lambda^*_{ij}}^2}} .$$

The above reparametrization facilitates in defining posterior distribution of $\boldsymbol {b_i}|\boldsymbol{z_i}, v_i$ as given by the following proposition.
\begin{proposition} \label{eqn:bprop}
Given the settings in \eqref{eqn:zi}, the conditional density
function of $\boldsymbol{b_i}|\boldsymbol{z_i}, v_i$ is specified by
\begin{equation} \label{eqn:bcon} \boldsymbol{b_i}|\boldsymbol{z_i}, v_i \sim N_q( \boldsymbol{\tau_i^2 D_i^\intercal \Psi_i^{-1}} (\boldsymbol{z_i -\Sigma_i^{1/2}\delta_i^* }v_i), \boldsymbol{\tau_i^2}) \end{equation}
where \[\boldsymbol{\tau^2_i}  = (\boldsymbol{\Omega^{-1}_{b} + D_i^\intercal \Psi_i^{-1}D_i})^{-1},
\quad \boldsymbol{\Psi_i} = \boldsymbol{\Sigma_i^{1/2}}(\boldsymbol{I-\delta_i^*\delta_i^{*T}})\boldsymbol{\Sigma_i^{1/2}}. \] Moreover,
\begin{equation} \label{eqn:bzv} \boldsymbol{b_i|z_i} \sim \SND_q \bigg(\boldsymbol{\tau^2_i D_i^\intercal\Psi_i^{-1}z_i , \tau_i^2+d_id_i^\intercal , \lambda_{b_i}}\bigg) \end{equation}
where 
\[ \boldsymbol{d_i} = \boldsymbol{\tau_i^2 D_i^\intercal \Psi_i^{-1}\Sigma_i^{1/2}\delta_i^*}, \quad \quad
\boldsymbol{\lambda_{b_i}} = -\frac{(\boldsymbol{ D_i^\intercal \Psi_i^{-1}\Sigma_i^{1/2} \delta_i^*})^\intercal(\boldsymbol{\tau_i^2 + d_i^\intercal d_i})^{1/2}}{\sqrt{1 +\boldsymbol{d_i}^\intercal(\boldsymbol{\tau_i^2})^{-1}\boldsymbol{d_i}}}. \]
\end{proposition}
\noindent Note that $\boldsymbol{\lambda_{b_i}}$ in \eqref{eqn:bzv} is completely specified, therefore, it does not increase the dimension of estimable vector of parameters. The proof of Proposition \ref{eqn:bprop}
is essentially based on Bayes' Theorem where % as evident from 
% \[f_{\boldsymbol{b_i}|\boldsymbol{z_i}, v_i} = \frac{f_{\boldsymbol{z_i}|\boldsymbol{b_i}, v_i}f_{\boldsymbol{b}_i}}{\int f_{\boldsymbol{z_i}|\boldsymbol{b_i}, v_i} f_{\boldsymbol{b}_i}d\boldsymbol{b}_i} ,\]
% where 
\begin{equation} \label{eqn:bayesz}
f_{\boldsymbol{z_i}| v_i} = \int f_{\boldsymbol{z_i}|\boldsymbol{b_i}, v_i} f_{\boldsymbol{b}_i}d\boldsymbol{b}_i = \Phi_{n_i}(\boldsymbol{\Sigma_i^{1/2}\delta_i^*}v_i,\boldsymbol{\Psi_i + D_i\Omega_b D_i^\intercal} ).
\end{equation}

Under general regularity conditions and by \eqref{eqn:comf}, the complete conditional log-likelihood is
\begin{equation}\label{eqn:likeli} \ell(\boldsymbol{\theta}|\boldsymbol{y,x, b}) =\sum_{i=1}^m \ell_i(\boldsymbol{\theta_i}|\boldsymbol{y_i,x_i, b_i}), \end{equation}
where by the hierarchical representation in \eqref{eqn:multiz} and \eqref{eqn:zi}
\begin{equation} \begin{aligned} \label{eqn:likelifinal} 
\ell_i(\boldsymbol{\theta_i}|\boldsymbol{y_i,x_i, b_i}) & \propto - \frac{1}{2} \log|\boldsymbol{\Psi_i}| \\ 
&- \frac{1}{2}(\boldsymbol{z_i -D_ib_i - \Sigma_i^{1/2}\delta_i^*}v_i)^\intercal\boldsymbol{\Psi_i}^{-1}(\boldsymbol{z_i -D_ib_i-\Sigma_i^{1/2}\delta_i^*}v_i)\\
& -\frac{1}{2} \sum_{j=1}^{n_i}\log (1- \delta_{ij}^2)  - \frac{1}{2}\sum_{j=1}^{n_i} \frac{(z_{ij} - (\boldsymbol{D_ib_i})_j -\delta_{ij}v_i)^2}{(1-\delta_{ij}^2)} \\
& + \sum_{j=1}^{n_i} \log f(y_{ij}|\theta_{ij}), \\
\end{aligned}\end{equation}
Given that $\boldsymbol{y}= (\boldsymbol{y}_{1}, \dots, \boldsymbol{y}_{m})$, $\boldsymbol{x}= (\boldsymbol{x}_{1}, \dots, \boldsymbol{x}_{m})$, $\boldsymbol{b}= (b_{1}, \dots, b_{m})$, and $|\boldsymbol{\Psi_i}|$ denotes the determinant of \(\boldsymbol{\Psi_i}\).

\subsection{Autoregressive correlation matrix}
To characterize the covariance matrix in a plausible manner, one
needs to take in to account different sources of random variation
within observations. Under the multiple observations per unit settings, these sources generally fall into three categories: measurement error, random effect, and serial correlation. The first source is controlled during the fitting process. The random effect source of variation is accounted for within the model as a random intercept $\boldsymbol{b_i}$. Therefore, we would only consider integrating the serial correlation source of variation, and as noted earlier the covariance matrix $\boldsymbol{\Sigma_i}$ presented in \eqref{eqn:yi} is modeled as a function of time and a dispersion variable $\xi_i$. Assuming a homogeneous variance within units, ($\sigma_i^2$), the
correlations amongst each unit observations ($\boldsymbol{Y_i}$)  are determined by the autocorrelation function $\rho_i(.)$  as
\begin{equation} \textsf{Cov}(Y_{ij} ,Y_{ik}) = \sigma_i^2 \rho_i(|t_{ij} - t_{ik}|). \end{equation}

The simplest form to express the serial correlation above is to
assume an explicit dependence of the current observation $Y_{ij}$ on
previous observations $Y_{i(j-1)}, \dots, Y_{i1}$, which could be modeled
using $n$-th order autoregressive model. For example,
considering a first order autoregressive model as
\begin{equation}y_{ij} = \alpha_i y_{i(j-1)} +\epsilon_{ij}, \quad \epsilon_{ij} \stackrel{iid}{\sim} N(0,\zeta) .\end{equation}
Note that it would be difficult to give an explicit interpretation of
the $\alpha$ parameter if the measurements are not equally spaced
in time or when times of measurements are not common to all units.
One way of solving this issue is to implement an exponential
autocorrelation function $\rho(.)$, where
\begin{equation}\label{eqn:varcovar} \textsf{Cov}( Y_{ij}, Y_{ik})  = \sigma_i^2 e^{ -\xi_i |t_{ij} - t_{ik}|}. \end{equation}
The correlation between two response variables is then
\begin{equation}\label{eqn:corr} \textsf{Corr}( Y_{ij}, Y_{ik})  = e^{ -\xi_i |t_{ij} - t_{ik}|}. \end{equation}
This correlation structure is used to construct the correlation coefficient matrix $\boldsymbol{\Sigma_i} = \boldsymbol{\Sigma_i}(\xi_i, \boldsymbol{t_i})$ in the copula structure and likelihood.
\section{Monte Carlo based EM algorithm}

The expectation-maximization (EM) algorithm (\cite{dempster}) is an iterative approach for obtaining the maximum likelihood estimates. It consists of two steps from which the name is derived; an expectation (E-step) and a maximization step (M-step). Typically the likelihood of interest involved a set of observed data $x$ and unobserved latent data $u$, where the conditional distribution of $u$ given $x$ is known. At iteration $r$, the E-step computes the expectation of the log-likelihood function with respect to the conditional distribution  $u|x, \theta^{(r)}$. The M-step computes a new set of (provisional) parameter estimates $\theta^{(r+1)}$ that maximize the expectation of the earlier E-step. Those two steps alternate to find a set of parameters that maximize the likelihood function. Let $\ell(\theta | u,x)$ be the log-likelihood, then, for $r=1,2\dots$ the alternating steps are as follows: 
\begin{itemize}
\item E-step: compute $Q(\theta |\theta^{(r)})=
E_{u|x,\theta^{(r)}}[\ell(\theta|u,x)]$; \item M-step: find
$\theta^{(r+1)} = \arg\max_\theta Q(\theta |\theta^{(r)})$.
\end{itemize}

Under certain regularity conditions discussed in \cite{Wu83}, the log-likelihood function converges to a local or global maximum.

The earliest detailed explanation and naming of the EM algorithm was published by \cite{dempster}, where they generalized earlier attempts by \cite{Sundberg1974}, and sketched a convergence analysis for a wider class of problems. \cite{Rubin93} studies computational difficulties encountered in the M-step, where they proposed smaller maximization steps over the parameter space. They argued that instead of maximizing the whole set of parameters one can maximize in a sequential manner a subset of parameters independently, while the other subset is held fixed. Such modification is called a constrained maximization step (CM). A second important advancement to the EM algorithm was proposed by \cite{MCEM}, and is called the Monte Carlo (MC) EM algorithm. By applying the law of large numbers on the E-step above, one can approximate $Q(\boldsymbol{\theta} |x,\boldsymbol{\theta}^{(r)})$ as
\begin{equation} \label{eqn:qulabel}Q(\boldsymbol{\theta} |\boldsymbol{\theta}^{(r)}) = R^{-1}
\sum_{t=1}^R \ell(\boldsymbol{\theta}|u^{(t)},x),\end{equation}
where $R$ is relatively a large sample size. 

In relation to the results discussed in earlier sections, the unobserved latent random variable is $\boldsymbol b_i$, where its conditional distribution  $\boldsymbol{b_i}|\boldsymbol{z_i,\theta_i}$ is found to be a skew-normal as illustrated in Proposition \eqref{eqn:bprop}. Therefore, let $\boldsymbol{\theta}^{(r)}$ be a vector of parameter estimates in the $r$-th iteration, then the two MC-EM steps are
\begin{itemize}
\item MC E-step: for the $i$-th unit at $(r+1)$ EM iteration,
\begin{equation} \label{eqn:QMCEM} \begin{aligned} Q_i(\boldsymbol{\theta}|\boldsymbol{\theta}^{(r)} ) &= E_{\boldsymbol{b_i}|\boldsymbol{z_i,\theta^{(r)}}}[ \ell_i(\boldsymbol{\theta} |\boldsymbol{x_i,y_i, b_i})] \\
& \cong R_i^{-1} \sum_{j=1}^{R_i} \ell_i(\boldsymbol{\theta} |\boldsymbol{x_i,y_i, b_i^{(j)}}), \end{aligned}
\end{equation}
and
\[  Q(\boldsymbol{\theta}|\boldsymbol{\theta}^{(r)} ) = \sum_{i=1}^m  Q_i(\boldsymbol{\theta}|\boldsymbol{\theta}^{(r)}),  \]
where $\boldsymbol{b}_i^{(j)}$ is the $j$-th draw generated from the
distribution of $\boldsymbol{b_i}|\boldsymbol{z_i,\theta^{(r)}}$, $R_i$ is the number of replication on the $i$-th unit. \item M-step: solving
the score equation
    \[ \frac{\partial}{\partial \theta} Q(\boldsymbol{\theta}|\boldsymbol{\theta}^{(r)} )  = 0.\]
\end{itemize}

It is important to mention the work of \cite{Wu83}, which outlined a list of conditions ensuring the convergence of the EM algorithm. Conditions as the boundedness of the log-likelihood, compactness of the parameter space and the continuity of the expectation in the E-step with respect to the estimated parameter. The log-likelihood of the proposed model in \eqref{eqn:likelifinal} involves a term of the form $\log(|\Psi|)$, which could reach infinity and compromise the convergence of the EM algorithm. To follow \cite{Wu83} conditions, heuristic methods of initiating the algorithm from different starting points is enforced in the MC-EM algorithm used in this paper. Similar heuristic methods were successfully used by \cite{Vale05}. The following sections illustrate some numerical and real data results of the proposed model and algorithm.

\subsection{An M-step for an exponential response} \label{sec:EMcases}
This subsection derives the likelihood and its partial derivatives when the response variable $Y_{ij}|x_{ij}, b_i$ follows an exponential distribution with mean function $\eta_{ij} = \exp(x_{ij}\beta +b_i)$, and density $f(y_{ij} |\eta_{ij})  = \eta^{-1}_{ij} \exp(- y_{ij} \eta^{-1}_{ij})$.

From \eqref{eqn:likelifinal} the unit log-likelihood is
\[ \begin{aligned} 
\ell_i(\boldsymbol{\theta_i}|\boldsymbol{y_i,x_i}, b_i) & \propto - \frac{1}{2} \log|\boldsymbol{\Psi_i}| \\ 
&- \frac{1}{2}(\boldsymbol{z_i} - b_i - \boldsymbol{\Sigma_i^{1/2}\delta_i^*}v_i)^\intercal\boldsymbol{\Psi_i}^{-1}(\boldsymbol{z_i} -b_i- \boldsymbol{\Sigma_i^{1/2}\delta_i^*}v_i)\\
&-\frac{1}{2} \sum_{j=1}^{n_i}\log (1- \delta_{ij}^2)  - \frac{1}{2}\sum_{j=1}^{n_i} \frac{(z_{ij} - b_i -\delta_{ij}v_i)^2}{(1-\delta_{ij}^2)} \\
&- \sum_{j=1}^{n_i}\{ y_{ij}e^{-x_{ij}\beta -b_i} +x_{ij}\beta + b_i\},\end{aligned}\]
where parameters are as defined in \eqref{eqn:likelifinal}. Therefore, the marginal partial derivatives become
\[ \frac{\partial}{\partial \beta}\ell_i(\boldsymbol{\theta_i}|\boldsymbol{y_i,x_i}, b_i) = \sum_{ j= 1}^{n_i} x_{ij} \{  y_{ij}e^{-x_{ij}\beta -b_i}  -1  \}\]
\[ \frac{\partial^2}{\partial \beta^2}\ell_i(\boldsymbol{\theta_i}|\boldsymbol{y_i,x_i}, b_i) = -\sum_{ j= 1}^{n_i} x^2_{ij} y_{ij}e^{-x_{ij}\beta -b_i}  \]
\[ I(\beta) = -\sum_{i=1}^mE\bigg (\frac{\partial^2}{\partial \beta^2} \ell_i(\boldsymbol{\theta_i}|\boldsymbol{y_i,x_i}, b_i)\bigg ) =\sum_{i=1}^m \sum_{j=1}^{n_i} x_{ij}^2 \]

\[\label{eqn:betahatexp}\hat{\beta} = - (\boldsymbol{X}^\intercal\boldsymbol{X})^{-1}\boldsymbol{X}^\intercal(\log(D^{-1}(\boldsymbol{Y})\mathbb{I}) +b_i\mathbb{I})  \]
where  $\mathbb{I} = (1, 1, \dots, 1)^\intercal$ and $D^{-1}$ is an inverse diagonal matrix. Similar results could be obtained using Gamma marginals with canonical link function.

\section{Simulation design and analysis}

To assess the efficiency of the proposed likelihood and model, a univariate and a bivariate model settings are used to infer parameters. Under both settings the  number of units is fixed to 200 and the number of observations $n_i$ is fixed to 5 for each unit. To generate the response variable \(\boldsymbol{Y}_i|b_i \), since the true parameters are known, we first generate the per-unit multivariate skew-normal variable \(\boldsymbol{Z}_i | b_i\) as in \eqref{eqn:zuni} with a specified skewness vector \(\boldsymbol{\lambda}\). Then, we use the inverse of the link of the marginal distributions of \(Z_{ij}\) and \(Y_{ij}\) defined in \eqref{eqn:martoscopula} to generate the per-unit multivariate response \(\boldsymbol{Y}_i|b_i \). The following two subsections discuss each model specific settings.

\subsection{Univariate model}\label{sec:univariate}
Here we use a model of a fixed intercept \(\alpha\) and a univariate random effect as 

\begin{equation} \boldsymbol{Y}_i|b_i \sim F_{n_i}(\eta(\alpha + b_i) , \boldsymbol{\Sigma}(\xi_i)), \end{equation}
where $F_{n_i}$ is a multivariate distribution from the exponential family with link function $\eta$, as in Section \eqref{sec:EMcases}.

The fixed and random effects coefficients are set as \(\alpha + b_i \sim N_1(3,2)\) such that $E[\alpha + b_i] = 3$. The time difference per observation within each unit is set to a unit difference, that is the elements of \(\boldsymbol{\Sigma}(\xi_i)\) are 

\begin{equation} \label{eq:uniTime} e^{-\xi_i|t_{ij} - t_{ik}|} = \left\{
  \begin{array}{l l}
    e^{-\xi_i|j-k|} & \quad \text{if } j \neq k\\
    1 & \quad \text{if } j =k ,\\
  \end{array} \right.
.\end{equation}
where \(\xi_i=\xi =0.2\). Finally, since we are simulating first the skew-normal variable \(\boldsymbol{Z}_i | b_i\) to get the response \(\boldsymbol{Y}_i|b_i \) we set the skewness vector \(\boldsymbol{\lambda}=(1, \dots, 1)\).

\subsection{Bivariate model}
This model investigates the convergence under extra variables, binary and categorical, which in some cases could represent a measurement deviation caused by certain events. We use a model structure similar to the one in Section 6 of \cite{Vale05} as

\begin{equation} \boldsymbol{Y}_i|b_i \sim F_{n_i}(\eta(\alpha+t_{ij}\beta_1 + \zeta_j \beta_2 + b_i ), \boldsymbol{\Sigma}(\xi_i)), \end{equation}

\noindent where \(\beta_1= 2\), \(\beta_2=1\) and \(t_{ij} = j-3\) for \(j=1, \dots,5\). A categorical variable \(\zeta_{j} = 1\) for \(i \leq 100\) and \(\zeta_{ij} =0\) otherwise. Similar to the univariate settings, we let \(\alpha + b_i \sim N_1(1, 4)\) such that \(E[\alpha + b_i]=1\) and \(Var[\alpha + b_i]=4\). The time difference per observation within each unit is set to a unit difference as in \eqref{eq:uniTime}, where \(\xi_i= \xi=0.2\), and the skewness vector \(\boldsymbol{\lambda}=(1, \dots, 1)\).

For each simulation of a 100, we set the initial estimates to $\boldsymbol{\beta}^{(0)} = 1$, $\lambda_i^{(0)}=0.5$, $Var[\alpha + b_i] =1$ and $\xi_i^{(0)} = 0.1$. Using the Monte Carlo EM algorithm, in each iteration we sample from $b^{(k)}|Z$, starting with $50$ samples per unit and gradually increasing until convergence.

\subsection{Exponential and gamma distributed response}
This section illustrates simulation results of the proposed copula-driven GLMM using the derived likelihood and the proposed MC-EM algorithm, and compares it numerically to the ordinary normal copula, where the skewness vector \(\boldsymbol \lambda\) is set to 0.

\begin{figure}[H]
\centerline{
\subfloat[A single replication]{\label{fig:expexpsingle} \includegraphics[scale=.45]{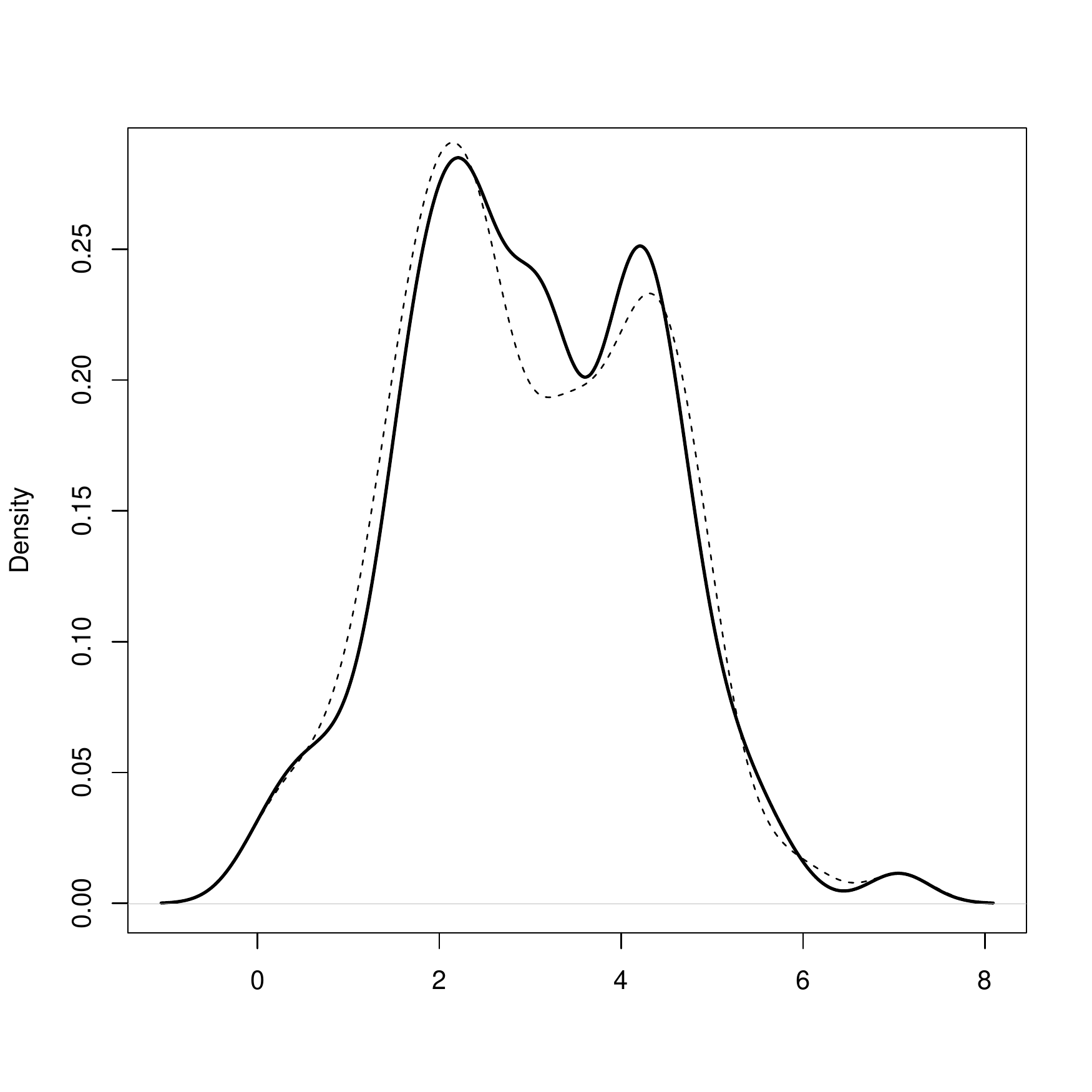}}
\subfloat[50 MC replications]{\label{fig:expexp100single} \includegraphics[scale=.45]{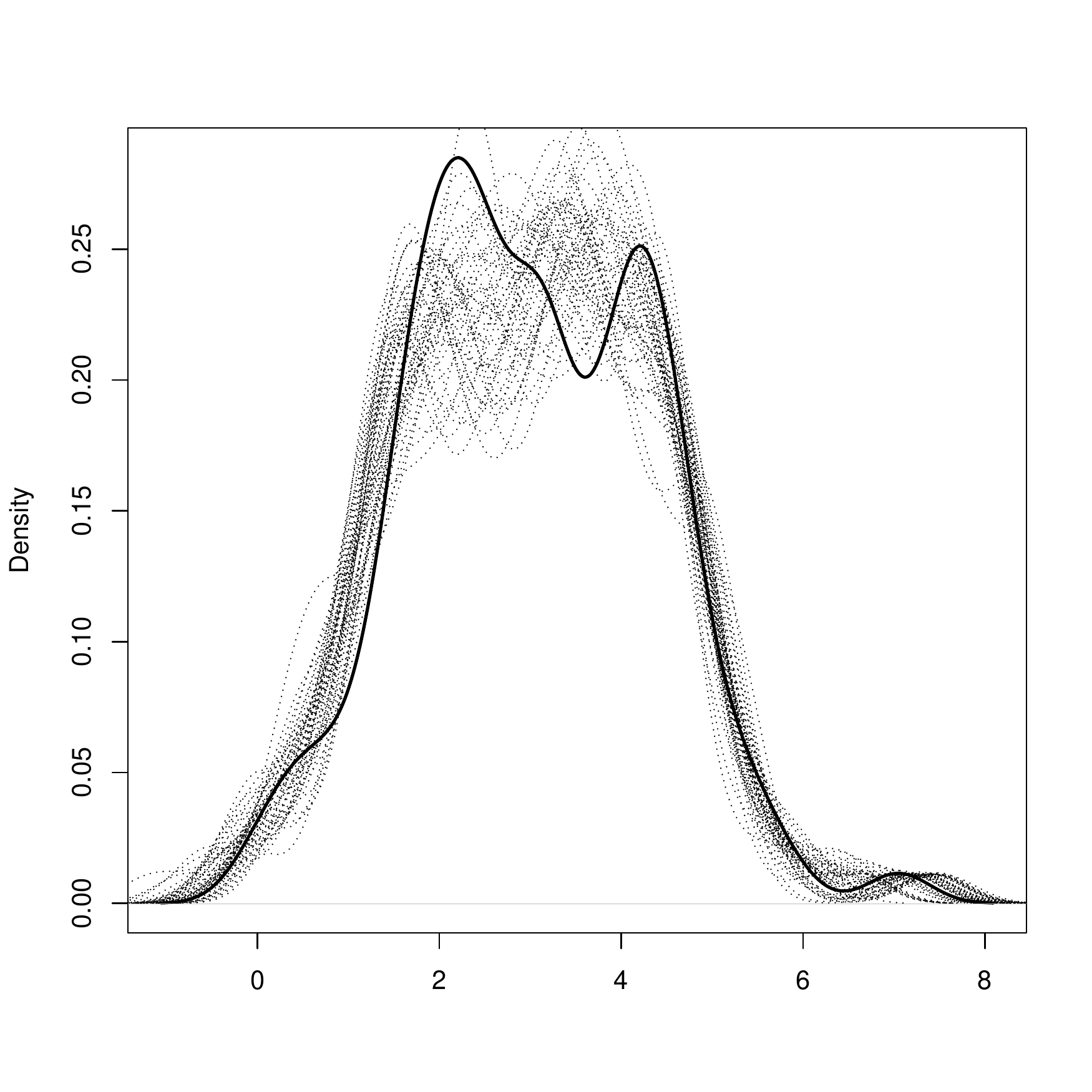}}
}
\caption{Univariate Settings with exponential marginals: the true and estimated density of the response variable $\boldsymbol Y$ on the log scale; in bold and dotted lines respectively.}
\label{fig:expexpuni}
\end{figure}

\begin{figure}[H]
\centerline{
\subfloat[A single replication]{\label{fig:expexpmulti1} \includegraphics[scale=.45]{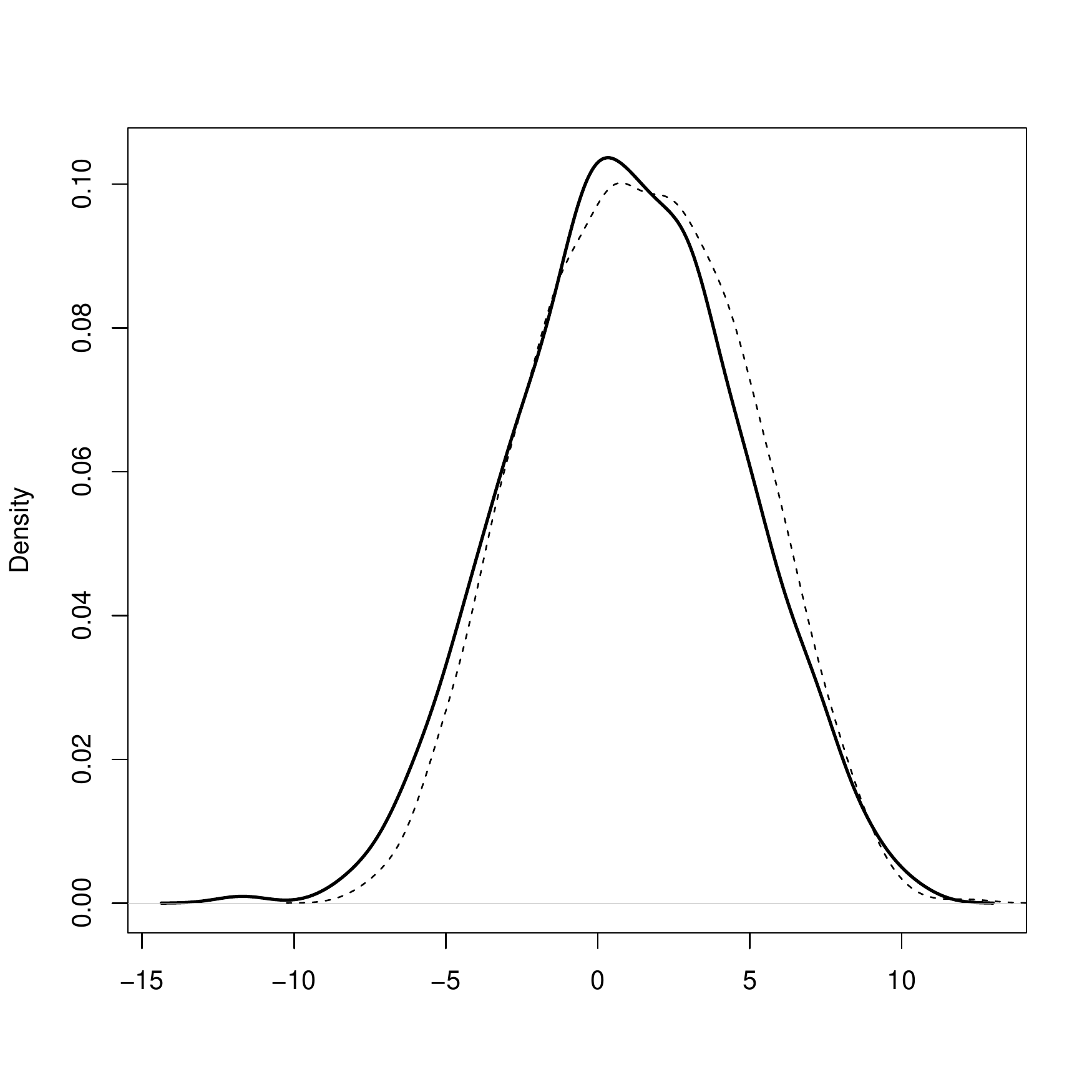}}
\subfloat[100 MC replications]{\label{fig:expexp100multi} \includegraphics[scale=.45]{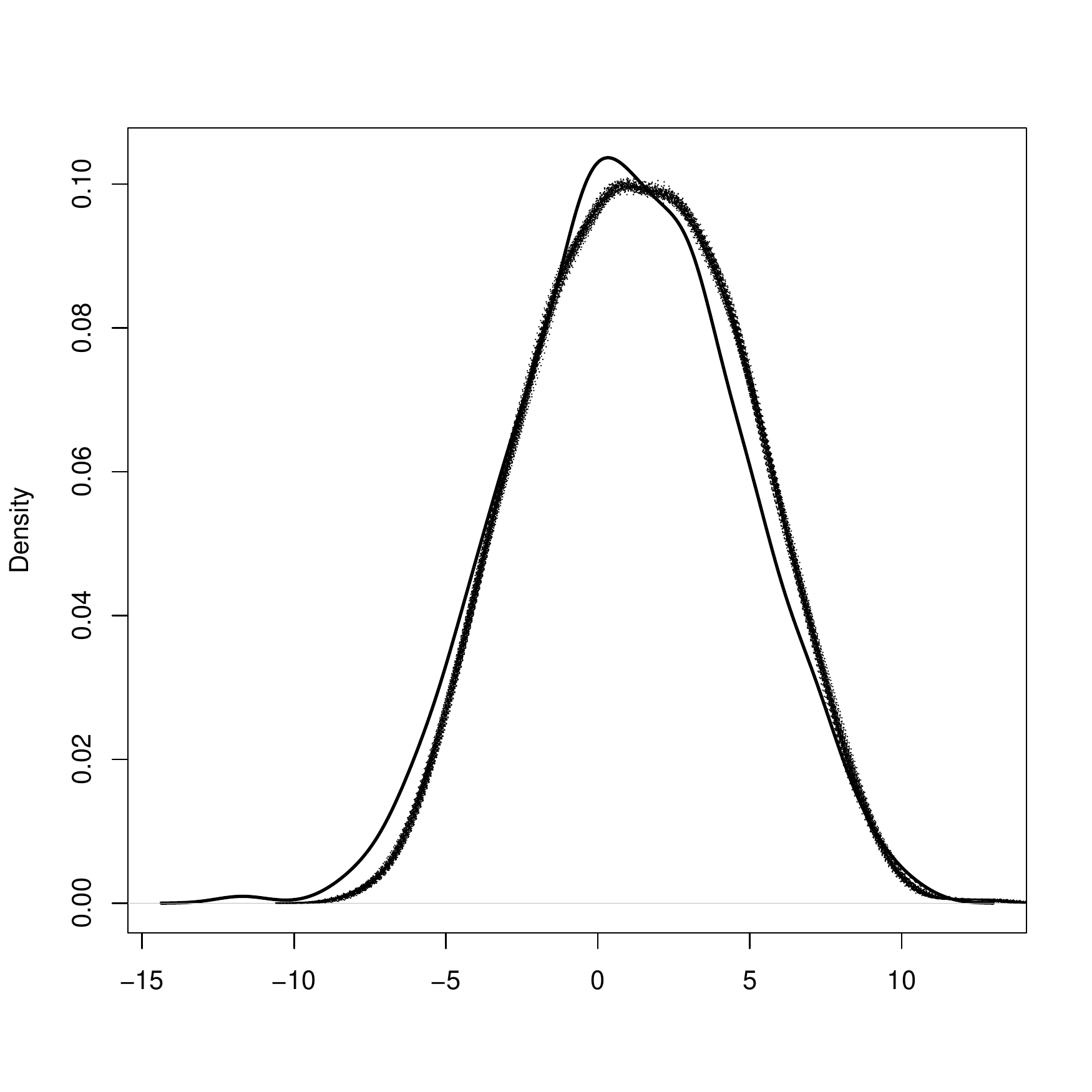}}
}
\caption{Bivariate settings with exponential marginals: the true and estimated density of the response variable $\boldsymbol Y$ on the log scale; in bold and dotted lines respectively.}
\label{fig:expexpmulti}
\end{figure}

The final missing piece of the likelihood in \eqref{eqn:likelifinal} is the specification of the marginal distribution of the response variable. Here we assume a response variable first from the exponential and then the gamma distribution with a log-link function. For each simulation a 100 Monte Carlo data sets are generated under the univariate and bivariate settings discussed in the previous subsections. Tables \ref{table:uniexpexp} and \ref{table:multiexpexp}, show the parameter estimates of the skew-normal on the left, and normal copula on the right, using exponential marginals, under the univariate and bivariate settings respectively. The MC Mean and MC SD represent the Monte Carlo mean and standard deviation.  MSE is the average standard error between Monte Carlo simulation and the true value of the parameter. EC represents the empirical coverage probability computed using Fisher information matrix assuming a 95\% confidence interval. Note that in the bivariate model we calculate the EC for \(\beta_1\) and \(\beta_2\) using a 95\% elliptical confidence interval. The \(\bar \lambda\) is the average skewness. Figures \ref{fig:expexpuni} and \ref{fig:expexpmulti} depict the convergence approximation graphically, under both models respectively for the skew-normal copula.\newline

\begin{table}[h]
	\centering
	\caption{Parameter estimation under the univariate settings with exponential marginals}
	\resizebox{13cm}{!}{
\begin{tabular}{lc|cccc|cccc}
	 \label{table:uniexpexp}
                      &                    & \multicolumn{4}{c}{Skew-normal copula}& \multicolumn{4}{c}{Normal copula}\\
	Parameters        & True value         & MC Mean    & MC SD    &  MSE  & EC   & MC Mean    & MC SD    &  MSE  & EC  \\
	\hline
	$\alpha$          & 3   & 2.889  & 0.0340 & 0.0135 & -    & 2.7440 & 0.0055 & 0.0655 & -   \\
	$E[\alpha + b]$   & 3   & 2.993  & 0.0121 & 0.0020 & 0.99 & 3.1623 & 0.0043 & 0.0264 & 0.00 \\
	$Var[\alpha + b]$ & 2   & 2.005  & 0.0959 & 0.0091 & 0.98 & 1.4732 & 0.0111 & 0.2776 & 0.00 \\
	$\xi$             & 0.2 & 0.2004 & 0.0062 & 0.0004 & -    & 0.1761 & 0.0016 & 0.0006 & -    \\
    $\bar \lambda$    & 1   & 1.205  & 0.0463 & 0.0441 & -    & -      & -      & -      & -    \\
	\end{tabular}}
\end{table}

\begin{table}[h]
\centering
\caption{Parameter estimation under the bivariate settings with exponential marginals}
\resizebox{13cm}{!}{
\begin{tabular}{lc|cccc|cccc}
 \label{table:multiexpexp}
                     &                    & \multicolumn{4}{c}{Skew-normal copula}& \multicolumn{4}{c}{Normal copula}\\
	Parameters       & True value         & MC Mean    & MC SD    &  MSE  & EC   & MC Mean    & MC SD    &  MSE  & EC  \\
	\hline
  $\alpha$          & 1    & 0.9274 & 0.0350 & 0.0065 & -    & 0.6691 & 0.0052 & 0.1095 & -    \\
  $\beta_1$         & 2    & 1.9781 & 0.0001 & 0.0005 & 0.99 & 1.9781 & 0.0001 & 0.0005 & 0.99 \\
  $\beta_2$         & 1    & 0.9492 & 0.0556 & 0.0056 & 0.99 & 0.9258 & 0.0061 & 0.0055 & 0.99 \\ 
  $E[\alpha + b]$   & 1    & 0.9488 & 0.0276 & 0.0034 & 0.66 & 1.1132 & 0.0046 & 0.0128 & 0.00 \\
  $Var[\alpha + b]$ & 4    & 4.1694 & 0.0955 & 0.0377 & 0.98 & 3.6689 & 0.0115 & 0.1098 & 0.00 \\ 
  $\xi$             & 0.20 & 0.2126 & 0.0049 & 0.0002 & -    & 0.1978 & 0.0013 & 0.0001 & -    \\ 
  $\bar \lambda$    & 1    & 0.9277 & 0.0273 & 0.0060 & -    & -      & -      & -      & -    \\ 
	\end{tabular}}
\end{table}

\begin{figure}[H]
\centerline{
\subfloat[A single replication]{\label{fig:gammaexp1multi} \includegraphics[scale=.45]{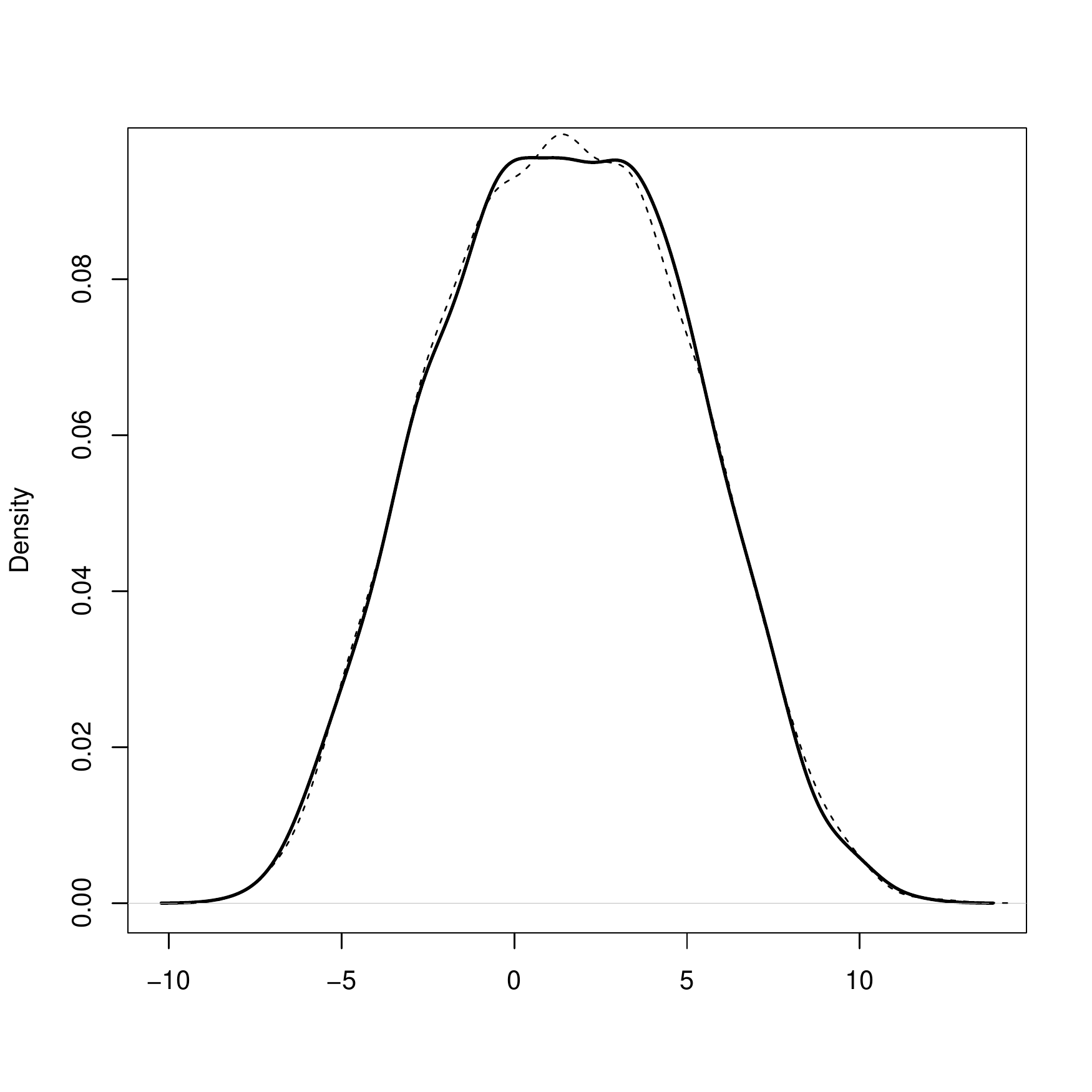}}
\subfloat[100 MC replications]{\label{fig:gammaexp100multi} \includegraphics[scale=.45]{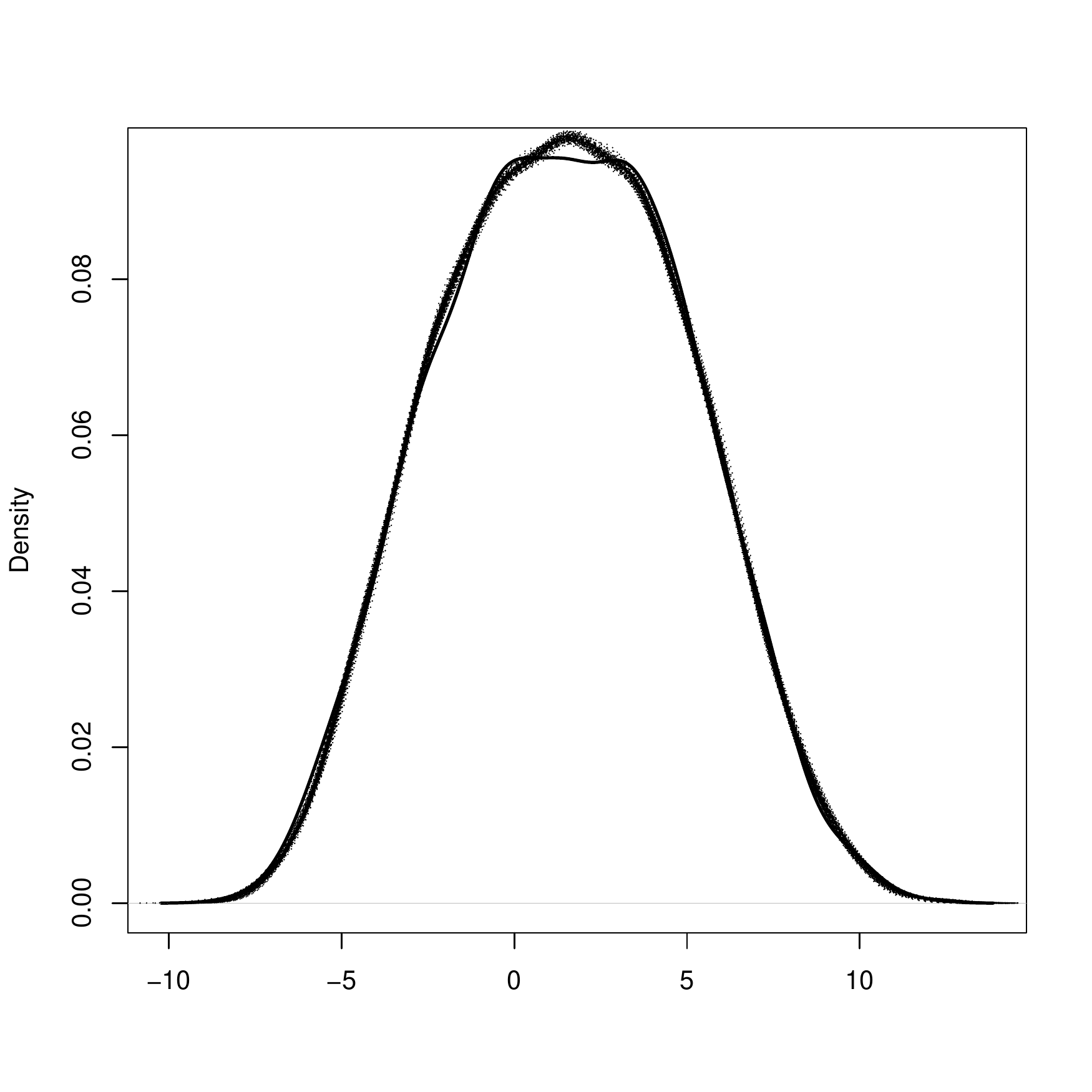}}
}
\caption{Bivariate settings with gamma marginals: the true and estimated density of the response variable $\boldsymbol Y$ on the log scale; in bold and dotted lines respectively.}
\label{fig:gammaexpmulti}
\end{figure}

Similarly, Figure \ref{fig:gammaexpmulti} and Table \ref{table:gammaexpmulti} show the simulation results of the bivariate model, while assuming gamma marginals. Table \ref{table:gammaexpmulti} also shows the estimated parameters when using the normal copula instead. The shape parameter of the gamma marginal is fixed to $k=3$ and a log-link function is used.

\begin{table}[!ht] 
	\centering
	\caption{Parameter estimation under the bivariate settings with gamma marginals}
\label{table:gammaexpmulti}
\resizebox{13cm}{!}{
	\begin{tabular}{lc|cccc|cccc}
       &                    & \multicolumn{4}{c}{Skew-normal copula}& \multicolumn{4}{c}{Normal copula}\\
	Parameters & True value         & MC Mean    & MC SD    &  MSE  & EC &  MC Mean    & MC SD    &  MSE  & EC  \\ \\
	\hline
  $\alpha$          & 1   & 0.8019 & 0.0417 & 0.0411 & -    & 0.8172 & 0.0066 & 0.0335 & -    \\
  $\beta_1$         & 2   & 2.0287 & 0.0001 & 0.0008 & 0.99 & 2.0288 & 0.0001 & 0.0008 & 0.99 \\ 
  $\beta_2$         & 1   & 0.9704 & 0.0611 & 0.0046 & 0.99 & 0.9332 & 0.0061 & 0.0045 & 0.99 \\
  $E[\alpha +b]$    & 1   & 0.9646 & 0.0363 & 0.0026 & 0.76 & 1.3660 & 0.0050 & 0.1339 & 0.00 \\
  $Var[\alpha + b]$ & 4   & 4.0288 & 0.0981 & 0.0104 & 0.98 & 3.4997 & 0.0123 & 0.2504 & 0.00 \\ 
  $\xi$             & 0.2 & 0.1987 & 0.0047 & 0.0001 & -    & 0.1815 & 0.0015 & 0.0003 & -    \\ 
  $\bar \lambda$    & 1   & 0.8998 & 0.0318 & 0.0110 & -    & -      & -      & -      & -    \\ 
\end{tabular}}
\end{table}

The results presented above suggest good inference results for the proposed model, since we are able to estimate the fixed parameters, the first and second moments of the random effects, and to some degree the autoregressive coefficient \(\xi\). Nevertheless, we intentionally fixed the number of observation per unit to 5, since it allows the use of a uniform \(\boldsymbol{\lambda}\) vector and an autoregressive parameter \(\xi\) for all units. In this sense, we can estimate the uniform parameters by drawing information from all observations. The reduction of the number of parameters is critical, since otherwise one has more parameters than observations. In our examples, using uniform autoregressive and skewness parameters, we only needed to estimate \(1 + 5\) parameters, while in general we have \(m + 5m\) parameters.

For the case when the normal copula is used, the estimation results of the fixed effects parameter is largely similar to the proposed model. This result is evident from \eqref{eqn:comf}, since the choice of the copula is independent from the likelihood of marginals. On the other hand, the estimation results for the random effect show systematic bias when compared to the results of the skew-normal model. This estimation bias arises from the fact that the skew-normal mean includes the skewness coefficient in its structure, thus it relates directly to the conditional distribution of \(\boldsymbol{b_i}|\boldsymbol{z_i}, v_i\), as seen in Proposition \ref{eqn:bprop}. In the case of the correlation parameter \(\xi\), the results are comparable with smaller differences in the bivariate setting, though a bit larger in the univariate setting, arguably due to the heavier influence of the random effects on the likelihood in the latter.

It is worth mentioning that the product form of the density in \eqref{eqn:comf} allowed the likelihood in  \eqref{eqn:likelifinal} to decomposed into three main parts. This in turn streamlined the estimation procedure of the fixed effects coefficient \(\boldsymbol{\beta}\) to the maximum-likelihood estimate when assuming independent marginal densities. One is then able to compute the information matrix analytically or by using methods as in \cite{louis1982finding} to obtain the observed information matrix. In this section we presented examples where the information matrix is readily available. Nevertheless, we find it to be much more complex to calculate the observed information matrix for the dispersion \(\xi\) and skewness \(\lambda\) variables, since it requires deriving the autoregressive correlation \(\boldsymbol{\Sigma}\) in \eqref{eqn:likelifinal} for the former and \(\boldsymbol{\Psi}\) for the latter. As a result, the coverage probabilities for both in the tables above are left blank.

The simulation was implemented in {\sf R} using mainly the packages {\sf sn} and {\sf mnormt}, which are both maintained by Adelchi Azzalini. The {\sf sn} package was used to sample from the skew-normal distribution and fit the skew-normal parameters, mainly  \(\boldsymbol{\widehat \Sigma}\) and \(\boldsymbol{\hat \lambda}\). Consequently, we estimate the dispersion parameter \(\xi\) by minimizing the \(L^2\) norm between the empirical estimate \(\boldsymbol{\widehat \Sigma}\) and the correlation matrix \(\boldsymbol{\widetilde{\Sigma}}(\xi)\) construed using \eqref{eq:uniTime}, as

\[ \hat \xi = \arg\min_{\xi >0}\bigg \{||\boldsymbol{\widehat \Sigma} - \boldsymbol{\widetilde{\Sigma}}(\xi)||_2 \bigg \}. \]

In respect to \(\hat \xi\) we then realign \(\boldsymbol{\widehat \Sigma}\) to \( \boldsymbol{\widetilde \Sigma}(\hat \xi)\). Likewise, one could also use the general-purpose optimization package {\sf optim} with {\sf L-BFGS-B} method with a lower bound of \(\tau>0\), less than an upper bound of \(\max  \{\delta^{\intercal} \delta\}\), to avoid singularities in computing the inverse of the matrix \(\boldsymbol{\Psi}\) in \eqref{eqn:zi} and \eqref{eqn:bcon}. Note that depending on the time measurement of observations \(\boldsymbol{t_i}\), the lower bound \(\tau\) cannot be very small, otherwise one will arrive at an all-ones matrix \(\boldsymbol{\widehat \Sigma}\). 

\section{An application}
As an illustration, we apply our methodology to the famous Framingham Heart Study that consists of longitudinal data for a wide set of cohorts. This data has been analyzed earlier in \cite{Zhang} and \cite{Vale05}. The primary objective is to model the change of cholesterol levels over time withing patients. The data provides cholesterol levels of 200 randomly selected patients, measured at the beginning of the study and every two years for a total of 10 years. However, we only use the first 3 observations per patient since it is the minimum number of visits seen in the data. The gender and age of those patients are also available. Since the normal linear mixed model analyzed by \cite{Zhang} is a particular case of GLMM, we apply our methodology to a simpler mixed model under more general distributional (copula based) setup. In view of the model proposed in Section 5, we consider the following model
\begin{equation}\label{eqn:chol} \boldsymbol{Y}_i \sim F_{n_i}(\alpha +\beta_1 sex_i +\beta_2 age_i  +\beta_3 \boldsymbol{t}_i + b_i, \boldsymbol{\Sigma}(\xi_i, \boldsymbol{t}_{i}) ), \end{equation}
where the $j$th component $y_{ij}$ of $Y_i$ is the cholesterol level at the $j$th time point for unit $i$ (the observations are normalize by a 100), $t_{ij}=(\text{time} -5)/10$ (time measured in years), $b_i$ is the unit specific random effect as in \eqref{eqn:multiz}, and the correlation coefficients are defined as
\begin{equation} Corr(Y_{ij}, Y_{ik}) = e^{-\xi\mid t_{ij} - t_i^* \mid},\end{equation}
\noindent where \(t_i^*\) is the time of the first visit.

\begin{figure}[H]
\centerline{
\subfloat[A single replication]{\label{fig:cholsingle} \includegraphics[scale=.45]{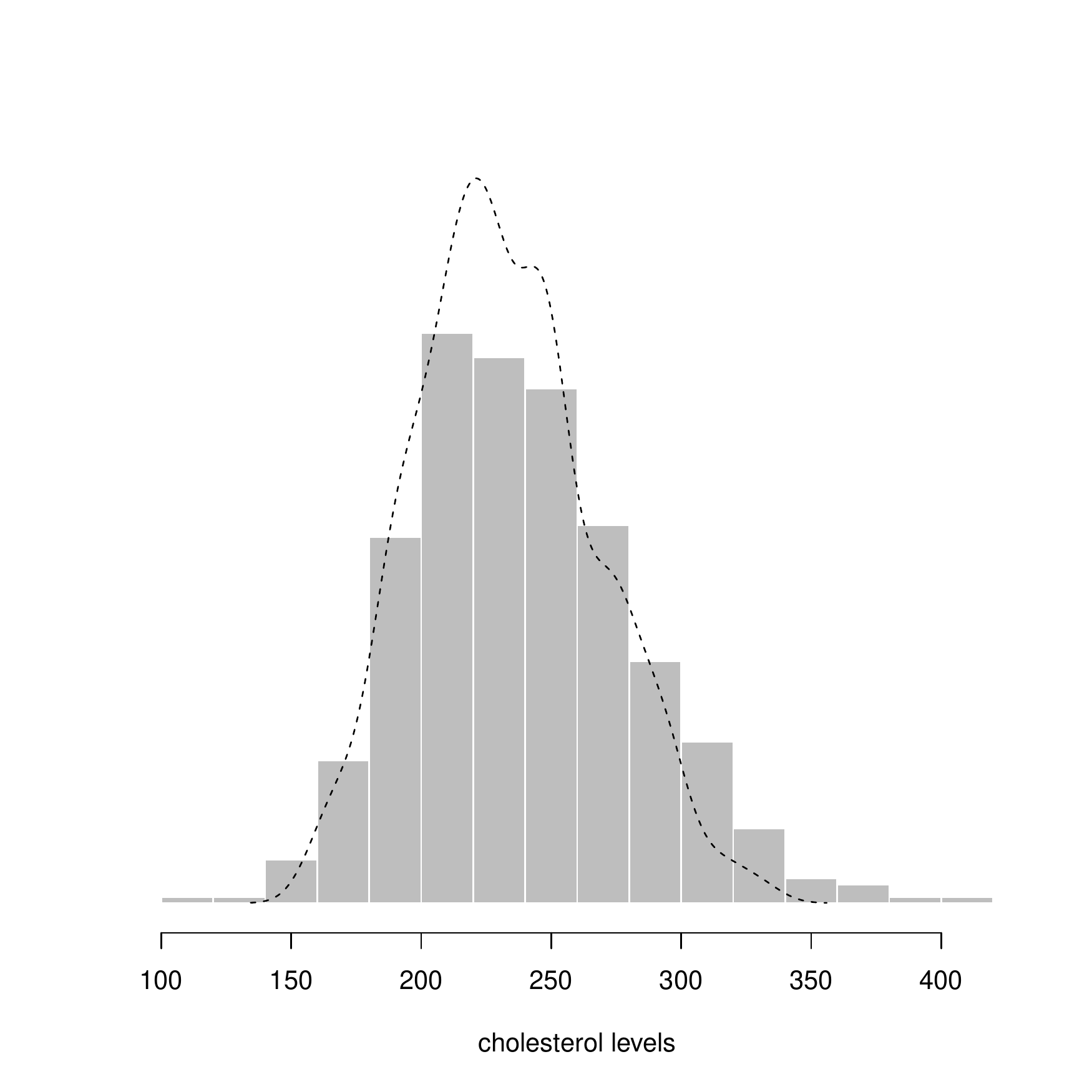}}\hspace{-1.7cm}
\subfloat[100 MC replications]{\label{fig:cholmulti} \includegraphics[scale=.45]{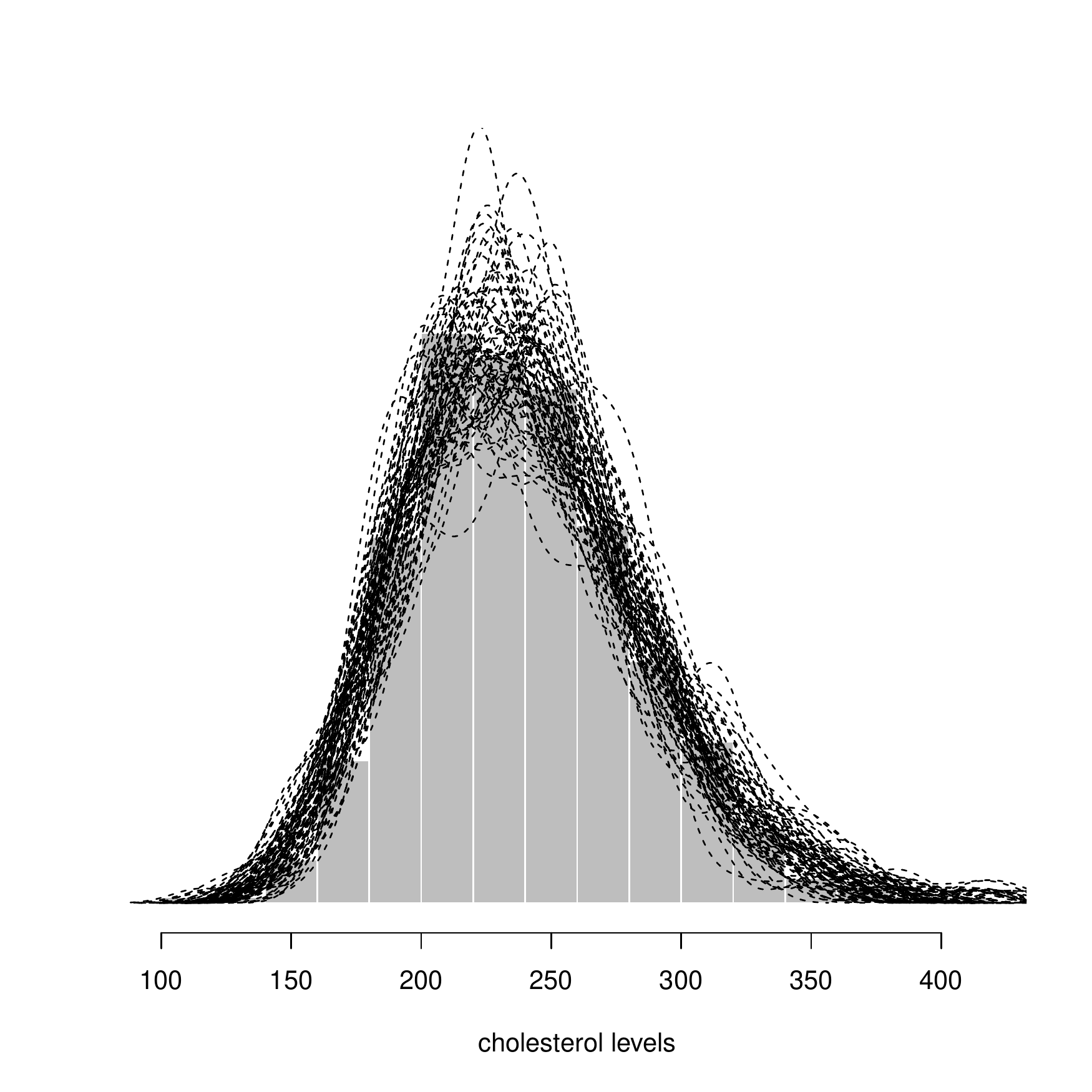}}
}
\caption{Fitting of Framingham Heart Study cholesterol data with model \eqref{eqn:chol} using a gamma marginals with a log-link function. The shape parameter is set to $k =3$. The solid lines are the fitted model, while the histogram shows the frequency distribution of cholesterol levels.}
\label{fig:chol}
\end{figure}

As in \eqref{eqn:martoscopula}, the modeling is performed with a gamma marginals and a log-link function. Figure \ref{fig:cholsingle} represents a histogram of cholesterol levels of the 200 randomly selected patients where dotted lines are the fitted model under the proposed settings. Figure \ref{fig:cholmulti} shows the same histogram with a 100 MC replications of $b_i$.

\begin{figure}[H]
\centerline{
\subfloat[Observed skew-normal densities]{\label{fig:empSN} \includegraphics[scale=.4]{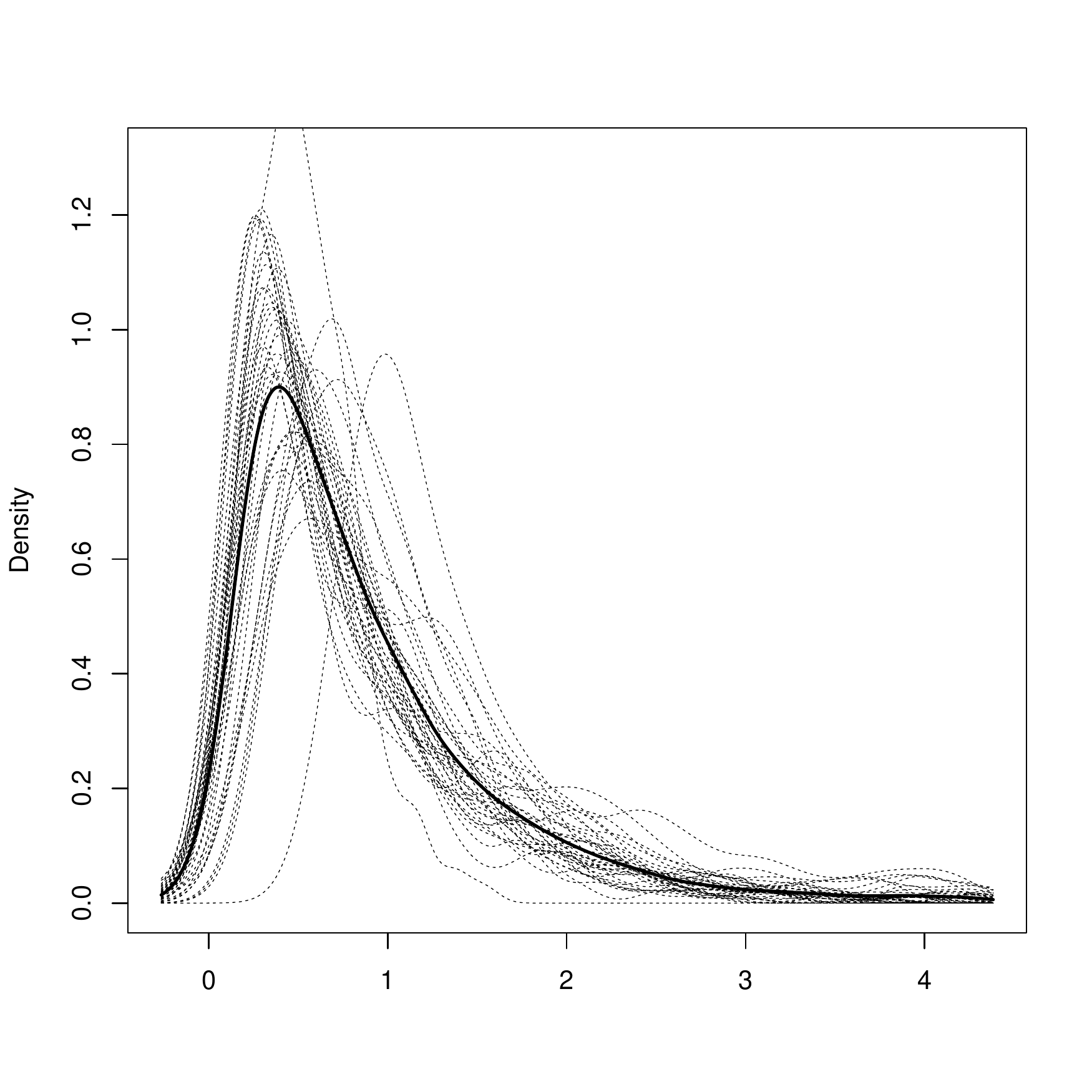}}\hspace{-0.5cm}
\subfloat[Average observed skew-normal versus fitted]{\label{fig:empvsreal} \includegraphics[scale=.4]{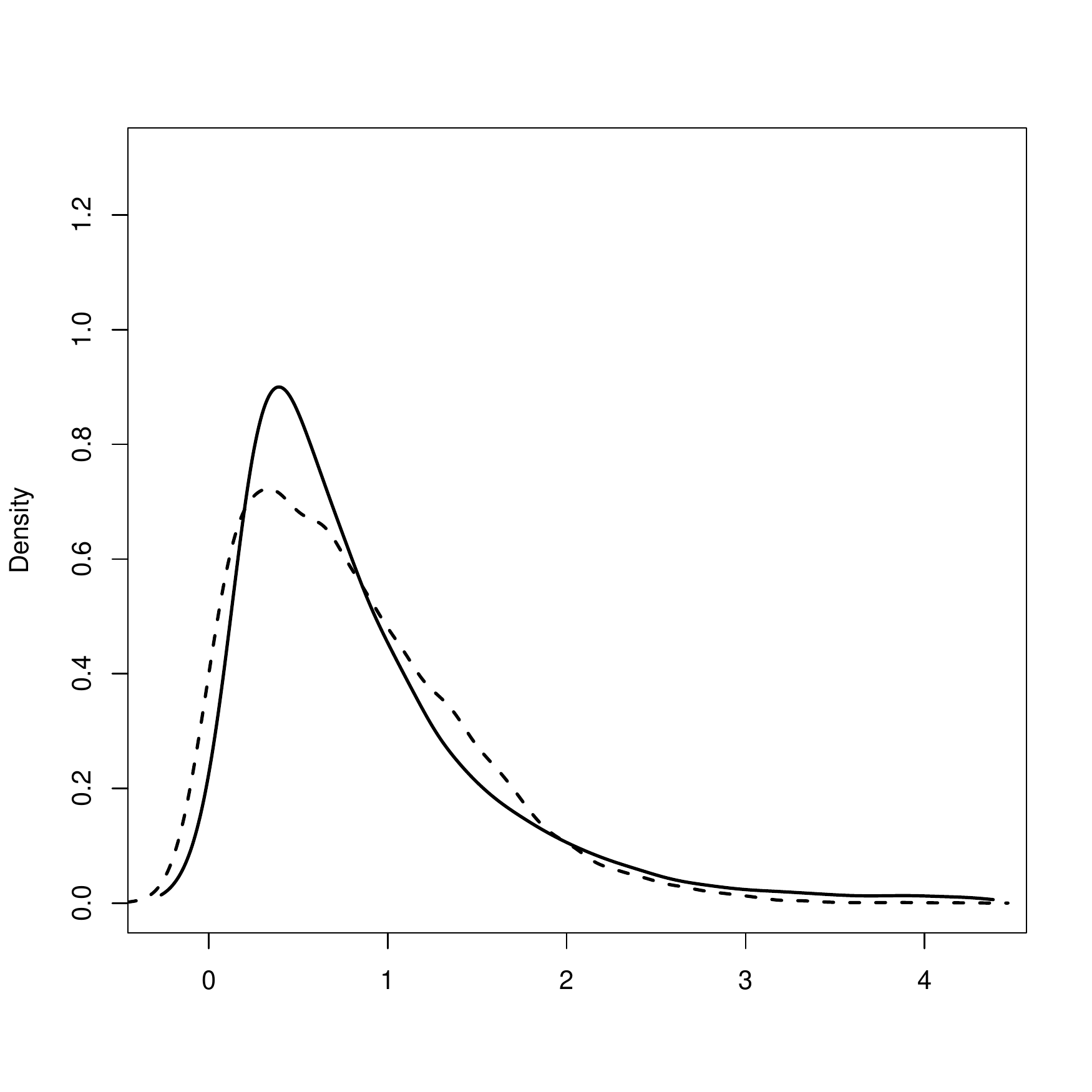}}
}
\caption{Figure on the left is the densities of centralized observed skew-normal from each MC-EM run. The figure on the right, in bold is the average density of the results in the left, while the dotted line is the density of a zero location skew-normal given the estimated parameters.  }
\label{fig:skewchol}
\end{figure}

Figure \ref{fig:empSN} represents the densities of the centralized observed skew-normal variable resulted from each of the 100 MC-EM runs, where the high positive skewness is evident. Figure \ref{fig:empvsreal} shows the density of the average centralized skew-normal variable in solid, versus the density of a zero location skew-normal generated using the fitted parameters.

\noindent Table \ref{table:chol1} presents the parameter estimates and standard
errors which are calculated as $SE(\theta_{\text{MLE}}) =
I(\theta_{\text{MLE}})^{-1/2}$, where $I$ is the Fisher Information
coefficient of the maximum likelihood estimate of parameter
$\theta$. From the table, the estimated value of the correlation
coefficient ($\xi$) is close to zero, this does not automatically imply that the proposed autoregressive correlation structure is not adequate. The normalization of the time variable \(\boldsymbol{t}_i\) affects the magnitude of \(\xi\). To see this better, the off-diagonal elements of the estimated correlation matrix \(\tilde \Sigma(\hat \xi)\) suggest a strong autoregressive structure in the data despite the low value of \(\hat \xi\).
\[\tilde \Sigma(\hat \xi) =  \begin{pmatrix}
  1 & 0.986 & 0.972 \\ 
   0.986 & 1 & 0.986 \\ 
   0.972 & 0.986 & 1 \\ 
 \end{pmatrix}.
\]
\noindent Moreover, \(\beta_2\) and \(\beta_3\)
estimates are close to zero, suggesting that patients age or time of observations are not a
predictor of cholesterol levels. Both \(\beta_1\) and \(Var[\alpha + b]\) seem relatively significant, emphasizing the importance of the patients gender and the random effects coefficient. The average skewness variable \(\bar \lambda\) suggests a highly skewed copula, as also indicated in \ref{fig:empSN}. Nevertheless, given the number of observations, the model has many variables to estimate, which dampen the estimation accuracy. In this case, we are estimating 9 coefficients for around 200 observations.

\begin{table}[H]
\centering
\caption{Fitting of Framingham Heart Study cholesterol data with model \eqref{eqn:chol} using a gamma marginals with a log-link function, the shape parameter $k =3$.}
\begin{tabular}{l*{6}{c}r}
\label{table:chol1}
Parameters              & Estimate & SE   \\
\hline
 \(\alpha\)         & 0.6394 & - \\ 
\(\beta_1\)         & 0.0764 & 0.0912 \\ 
\(\beta_2\)         & 0.0020 & 0.0055 \\ 
\(\beta_3\)         & -0.0012 & 0.0904 \\ 
\(E[\alpha + b]\)   & 0.8019 & 0.2914 \\ 
\(Var[\alpha + b]\) & 0.3173 & 0.0276 \\ 
\(\xi\)              & 0.0241 & - \\
\(\bar\lambda\)     & 4.4426 &  - \\ 
\hline
-log-likelihood     &   -1324.4    \\
AIC                 &    3.627    \\
BIC                 &    41.576  \\
\end{tabular}
\end{table}

\noindent \cite{Vale05} fitted the Framingham Heart Study cholesterol data under a mixture of Gaussian and skew-normal distributions for the random effects and residuals. In their model they used a bivariate random effect, while the presented model in \eqref{eqn:chol} uses a univariate random effect. Moreover, \cite{Vale05} used a linear mixed model formulation which differs from the copula formulation used here. For these differences, the average mean squared error of \cite{Vale05} surpasses the fit of the proposed model. Nonetheless, we believe the copula formulation allows more flexibility in modelling the response variable given a robust estimation procedure. In addition, this is the first step to estimate mixed models via a skew-normal copula, and future research is required to determine better fits, and most importantly, to integrate a random effects design matrix, and improve the estimation of the skewness and autoregressive variables.

\section{Discussion and future work}
The current investigation is based on the development of a copula-driven GLMM, where the focus was on modeling the marginals in lieu of the joint distribution. Oftentimes marginal distributions from the exponential family do not necessarily lead to a multivariate distribution of the same form. Nonetheless, we feel that copula based general multivariate distributions may be of more interest to applied statisticians. Our proposal intended to illustrate such a typical situations. 

In regard to the methodology, the MCEM seems to be appealing, though computationally expensive. We feel that estimation accuracy of the proposed model is pigged to the theoretical limitation of the EM algorithm, specially in large dimensions. Oftentimes, the MCEM algorithm converged to local maximums, and we feel that a post-EM optimization procedure, such as gradient descend, might improve the fit. One can also get rid of computational hassle to some extent by adopting a MCMC in the Bayesian paradigm.

In our subsequent investigations, we are planning to work with a Bayesian paradigm in a more broad setup. More importantly, we are planning to integrate a design matrix for the random effects to extend it beyond the univariate case. To improve the accuracy, we are attempting different optimization techniques. For computational convenience, an autoregressive structure was used to model the correlation matrix, which is not always applicable in real data, thus, we are planning to investigate more flexible correlation models. \newline

{\bf Acknowledgments:}\\
We would like to acknowledge the Associate Editor and all reviewers for their valuable comments.

%------------------------------
% Extra citation from referees.
\nocite{Meza12}\nocite{Peter09}\nocite{Walker14}\nocite{kaarik2015}

\bibliographystyle{plainnat}
\renewcommand{\bibname}{References}
\bibliography{references}

\begin{thebibliography}{23}
\providecommand{\natexlab}[1]{#1}
\providecommand{\url}[1]{\texttt{#1}}
\expandafter\ifx\csname urlstyle\endcsname\relax
  \providecommand{\doi}[1]{doi: #1}\else
  \providecommand{\doi}{doi: \begingroup \urlstyle{rm}\Url}\fi

\bibitem[Arellano-Valle and Genton(2005)]{Genton}
R.~Arellano-Valle and M.G. Genton.
\newblock Fundamental skew distributions.
\newblock \emph{Journal of Multivariate Analysis}, 96:\penalty0 93--116, 2005.

\bibitem[Arellano-Valle et~al.(2005)Arellano-Valle, Bolfarine, and
  Lachos]{Vale05}
R.~Arellano-Valle, H.~Bolfarine, and V.~Lachos.
\newblock Skew-normal linear mixed models.
\newblock \emph{Journal of Data Science}, 3:\penalty0 415--438, 2005.

\bibitem[Azzalini(1985)]{Azza85}
A.~Azzalini.
\newblock A class of distributions which includes the normal ones.
\newblock \emph{Scandinavian Journal of Statistics}, 12:\penalty0 171--178,
  1985.

\bibitem[Azzalini and Dalle-Valle(1996)]{AzzDalle96}
A.~Azzalini and A.~Dalle-Valle.
\newblock The multivariate skew-normal distribution.
\newblock \emph{Biometrika}, 83:\penalty0 715--726, 1996.

\bibitem[Azzalini(2013)]{azzalini2013skew}
Adelchi Azzalini.
\newblock \emph{The skew-normal and related families}, volume~3.
\newblock Cambridge University Press, 2013.

\bibitem[Dempster et~al.(1977)Dempster, Laird, and Rubin]{dempster}
A.P. Dempster, N.M. Laird, and D.B. Rubin.
\newblock Maximum likelihood from incomplete data via the {EM} algorithm.
\newblock \emph{Journal of the Royal Statistical Society}, 39\penalty0
  (1):\penalty0 1--38, 1977.

\bibitem[Fisher(1918)]{Fisher}
R.A. Fisher.
\newblock The correlation between relatives on the supposition of {M}endelian
  inheritance.
\newblock \emph{Transactions of the Royal Society of Edinburgh}, 52:\penalty0
  399--433, 1918.

\bibitem[Henze(1986)]{Henze86}
N.~Henze.
\newblock A probabilistic representation of the 'skew-normal' distribution.
\newblock \emph{Scandinavian Journal of Statistics}, 13\penalty0 (4):\penalty0
  271--275, 1986.
\newblock ISSN 03036898, 14679469.

\bibitem[K{\"a}{\"a}rik et~al.(2015)K{\"a}{\"a}rik, Selart, and
  K{\"a}{\"a}rik]{kaarik2015}
Meelis K{\"a}{\"a}rik, Anne Selart, and Ene K{\"a}{\"a}rik.
\newblock On parametrization of multivariate skew-normal distribution.
\newblock \emph{Communications in Statistics-Theory and Methods}, 44\penalty0
  (9):\penalty0 1869--1885, 2015.

\bibitem[Lambert and Vandenhende(2002)]{SIM:SIM1249}
P.~Lambert and F.~Vandenhende.
\newblock A copula-based model for multivariate non-normal longitudinal data:
  analysis of a dose titration safety study on a new antidepressant.
\newblock \emph{Statistics in Medicine}, 21\penalty0 (21):\penalty0 3197--3217,
  2002.

\bibitem[Landsman(2009)]{Landsman}
Z.~Landsman.
\newblock Elliptical families and copulas: tilting and premium; capital
  allocation.
\newblock \emph{Scandinavian Actuarial Journal}, 2009\penalty0 (2):\penalty0
  85--103, 2009.

\bibitem[Louis(1982)]{louis1982finding}
Thomas~A. Louis.
\newblock Finding the observed information matrix when using the em algorithm.
\newblock \emph{Journal of the Royal Statistical Society. Series B
  (Methodological)}, pages 226--233, 1982.

\bibitem[Meng and Rubin(1993)]{Rubin93}
X.L. Meng and D.B. Rubin.
\newblock Maximum likelihood estimation via the {ECM} algorithm: A general
  framework.
\newblock \emph{Biometrika}, 80:\penalty0 67--278, 1993.

\bibitem[Meza et~al.(2012)Meza, Osorio, and De~la Cruz]{Meza12}
C.~Meza, F.~Osorio, and R.~De~la Cruz.
\newblock Estimation in nonlinear mixed-effects models using heavy-tailed
  distributions.
\newblock \emph{Statistics and Computing}, 22\penalty0 (1):\penalty0 121--139,
  2012.

\bibitem[Newton and Zhang(1999)]{Newton}
M.A. Newton and Y.~Zhang.
\newblock A recursive algorithm for non-parametric analysis with missing data.
\newblock \emph{Biometrica}, 1999.

\bibitem[Petersson et~al.(2009)Petersson, Hanze, Savic, and Karlsson]{Peter09}
K.J.F. Petersson, E.~Hanze, R.M. Savic, and M.O. Karlsson.
\newblock Semiparametric distributions with estimated shape parameters.
\newblock \emph{Pharmaceutical Research}, 26\penalty0 (9):\penalty0 2174--2185,
  2009.

\bibitem[Sundberg(1974)]{Sundberg1974}
R.~Sundberg.
\newblock Maximum likelihood theory for incomplete data from an exponential
  family.
\newblock \emph{Scandinavian Journal of Statistics}, 1\penalty0 (2):\penalty0
  pp. 49--58, 1974.

\bibitem[Tao et~al.(1999)Tao, Aptal, Yandell, and Newton]{Tao}
H.~Tao, M.~Aptal, B.S. Yandell, and M.A. Newton.
\newblock An estimation method for the semi-parametric mixed effects model.
\newblock \emph{Biometrics}, 55:\penalty0 102--110, 1999.

\bibitem[Verbeke and Lesaffre(1996)]{Verbeke}
G.~Verbeke and E.~Lesaffre.
\newblock A linear mixed-effects model with heterogeneity in the random-effects
  population.
\newblock \emph{Journal of the American Statistical Association}, 91:\penalty0
  217--221, 1996.

\bibitem[Wei and Tanner(1990)]{MCEM}
G.C.G. Wei and M.A. Tanner.
\newblock A {M}onte {C}arlo implementation of the {EM} algorithm and the poor
  man’s data augmentation algorithms.
\newblock \emph{Journal of the American Statistical Association}, 85:\penalty0
  699--704, 1990.

\bibitem[Wu(1983)]{Wu83}
C.F.J. Wu.
\newblock On the convergence properties of the {EM} algorithm.
\newblock \emph{The Annals of Statistics}, 11\penalty0 (1):\penalty0 95--103,
  1983.

\bibitem[Wu et~al.(2014)Wu, Wang, and Walker]{Walker14}
J.~Wu, X.~Wang, and S.G. Walker.
\newblock Bayesian nonparametric inference for a multivariate copula function.
\newblock \emph{Methodology and Computing in Applied Probability}, 16\penalty0
  (3):\penalty0 747--763, 2014.

\bibitem[Zhang and Davidian(2001)]{Zhang}
D.~Zhang and M.~Davidian.
\newblock Linear mixed models with flexible distributions of random effects for
  longitudinal data.
\newblock \emph{Biometrics}, 57:\penalty0 795--802, 2001.

\end{thebibliography}
\vspace{0.25in}
\footnotesize{ 

%\noindent Kaylan Das\\
}
 \end{document}